\documentclass{raa_onecolumn}
\usepackage{graphicx,times}
\usepackage{natbib}
\usepackage{amssymb,amsmath}
\bibpunct{(}{)}{;}{a}{}{,}

\long\def\Ignore#1{\relax}

\newcommand{\kms}{{\rm km~s^{-1}}}
\newcommand{\kpc}{{\rm kpc}}
\newcommand{\kmskpc}{\kms\kpc^{-1}}

\newcommand{\Ms}{M_\odot}
\newcommand{\degree}{\circ}
\newcommand{\omegap}{\Omega_{\rm p}}

\begin{document}

   \title{The Bar and Spiral Arms in the Milky Way: Structure and Kinematics}

 \volnopage{ {\bf 2020} Vol.\ {\bf 20} No. {\bf XX}, 000--000}
   \setcounter{page}{1}
%（沈俊太） （郑兴武）
\author{{Juntai Shen}
      \inst{1,2}
   \and {Xing-Wu Zheng}
      \inst{3}   }

   \institute{Department of Astronomy, School of Physics and Astronomy, Shanghai Jiao Tong University, 800 Dongchuan Road, Shanghai 200240, China; email: {\it jtshen@sjtu.edu.cn}\\
   \and
   Shanghai Key Laboratory for Particle Physics and Cosmology, Shanghai 200240, China \\
   \and
   Department of Astronomy, Nanjing University, Nanjing 210093, China; {\it xwzheng@nju.edu.cn}\\
\vs \no
   {\small Received 2020 September 16; accepted 2020 XXX}
}

\abstract{The Milky Way is a spiral galaxy with the Schechter characteristic luminosity $L_*$, thus an important anchor point of the Hubble sequence of all spiral galaxies. Yet the true appearance of the Milky Way has remained elusive for centuries. We review the current best understanding of the structure and kinematics of our home galaxy, and present an updated scientifically accurate visualization of the Milky Way structure with almost all components of the spiral arms, along with the COBE image in the solar perspective. The Milky Way contains a strong bar, four major spiral arms, and an additional arm segment (the Local arm) that may be longer than previously thought. The Galactic boxy bulge that we observe is mostly the peanut-shaped central bar viewed nearly end-on with a bar angle of $\sim 25-30^\degree$ from the Sun-Galactic center line. The bar transitions smoothly from a central peanut-shaped structure to an extended thin part that ends around $R\sim 5 \;\kpc$. The Galactic bulge/bar contains $\sim 30-40\%$ of the total stellar mass in the Galaxy. Dynamical modelling of both the stellar and gas kinematics yields a bar pattern rotation speed of  $\sim 35 - 40 \; \kmskpc$, corresponding to a bar rotation period of $\sim 160-180$~Myr. From a galaxy formation point of view, our Milky Way is probably a pure-disk galaxy with little room for a significant merger-made, ``classical'' spheroidal bulge, and we give a number of reasons why this is the case. }

   \authorrunning{{\it Shen \& Zheng}: The Bar and Spiral Arms in the Milky Way}            %author_head in even pages
   \titlerunning{{\it Shen \& Zheng}: The Bar and Spiral Arms in the Milky Way}  % title_head in odd pages
   \maketitle


%\tableofcontents

%________________________________________________ sections below
%
\section{Introduction}           %% first-level sections will be auto-capitalized
\label{sect:intro}

What the Milky Way looks like has long been a mystery. There are three main reasons why this problem has been unsolved for so long. One is that most of the visible stars lie in the Galactic thin disk. Our solar system lies nearly in the mid-plane of that disk, far from the center of the Milky Way, and from our vantage point we cannot distinguish structures in the Milky Way because of projection effects. The difficult situation that astronomers face is best illustrated by a famous Chinese classic poem by Su Dongpo in the Song dynasty: ``I don't see the true face of Mountain Lushan because I myself am on the mountain''.  Secondly, thick dust clouds block optical light from distant stars in the disk of the Milky Way. Thirdly, most stars are very distant. The stars in the closest major arm of the Milky Way, the Sagittarius-Carina arm, are about 1400 pc from us.
As a result, we may never take a real optical picture of the Milky Way, so instead we create models based on measuring distances to objects that trace Galactic structure. A successful Galactic model must also explain the motions of gas and stars in the gravitational potential of the Milky Way.
The challenge is to synthesize all the direct and indirect information to weave a complete picture of the Milky Way.

The most prominent features of the visible part of the Milky Way are the Galactic bar and spiral arms.  This review focuses mainly on the structure and kinematics of the Galactic bar/bulge and spiral arms, on which the famous Hubble classification scheme of spiral galaxies is based. Only with accurate information of Galactic bar/bulge and spiral arms can we pinpoint the exact location of the Milky Way in the Hubble sequence of spiral galaxies.
%The famous Hubble classification scheme (``tuning-fork'') of galaxies classifies spiral galaxies based on their properties of bulge, spiral structure, and barredness.

For nearly a hundred years, the story of building a picture of the Milky Way has been the story of finding good tracers of spiral structure and credible methods of measuring their distances. The global spiral structure of the Galaxy first became apparent through study of the 21-cm line of neutral hydrogen. \citet{Oort_etal_58} produced a map using the intensity and Doppler shift of the 21-cm line to extract kinematic distances under the assumption that hydrogen clouds are in pure circular rotation about the Galactic center. This map showed long arcs of gas that resembled the spiral features in optical images of external galaxies. By using kinematic and spectrophotometric methods to determine the distances of young OB stars and giant HII regions, \citet{geo_geo_76} constructed a picture of the Milky Way with four spiral arms. In the 2010s \citet{lumsde_etal_13} used their Red MSX Source survey to map the structure of the Milky Way using about 1650 massive young stellar objects (MYSOs) and HII regions. Their model of the Milky Way \citep{urquha_etal_14, cabrer_etal_07} also had four spiral arms. However, they did not delineate arm locations accurately because of the dispersion of MYSO properties. By counting the near-infrared and mid-infrared stars near the tangent point from the Spitzer/GLIMPSE (Galactic Legacy Mid-Plane Survey Extraordinaire) and COBE/DIRBE/ZSMA surveys of the Galactic plane, \citet{benjam_etal_05} and \citet{drimme_etal_00} reported that the Milky Way has only two major stellar arms, the Perseus arm and the Scutum-Centaurus arm. Using a combined tracer sample of Galactic HII regions, GMCs, and 6.7 GHz methanol masers, \citet{hou_han_14} outlined the spiral structure and found that models of three-arm and four-arm logarithmic spirals are able to connect most spiral tracers \citep[also][]{hou_etal_09}.

Such disagreements about the spiral structure of the Milky Way might be explained by the different tracers and different approaches used to determine their distances by different researchers. Kinematic and spectrophotometric methods to determine the distances of MYSOs and giant HII regions suffer considerable uncertainties because of inadequacies in the velocity-to-distance and luminosity-to-distance relationships \citep{burton_etal_92}. Total star counts may not allow good mapping of the spiral arms in the Milky Way, where there might be different contents of evolved stars and gas.
%(why? need to explain a bit).

Accurate distance measurement is crucial in resolving the disagreements among different groups.
Trigonometric parallax provides the most reliable distance determination for a stellar object, and has revolutionized the field in the last twenty years. It is a completely geometric method, independent of any assumptions or astrophysical models. Interstellar masers, such as those of water vapor ($\mathrm H_2O$) and methanol ($\mathrm CH_3OH$), are the most important signposts to high-mass star-forming regions (HMSFRs) along the spiral arms in the Milky Way. The microwave emission from such masers penetrates the dust and gas in the disk of the Galaxy and can be very bright, so that they can be detected over the entire Milky Way. Masers are very compact objects within complexes of size 20-30 pc \citep{rei_mor_81} and are ideal for precise parallax measurements.  The Bar and Spiral Structure Legacy (BeSSeL) survey aims to determine accurate distances of HMSFRs in the bar and spiral arms of the Milky Way by measuring the trigonometric parallaxes of the methanol and water masers with which they are associated. The spiral structure with unprecedented accuracy revealed by the BeSSeL survey is summarized in \S\ \ref{sect:spirals}.

Compared to the spiral arms, little was known about the structure of the Milky Way bar until relatively recently since the central bar is still much further away from the Sun than the nearby spiral arms. Before the 1990s the Milky Way was once considered as an unbarred galaxy to most of the astronomical community. The widely-held but erroneous belief that the Milky Way is unbarred puzzled many theorists since dynamically ``cold'' galactic disks were known to be violently unstable to large-scale instabilities that can result in a strong bar. Intriguingly, based on the unbarred ``fact'' of the Milky Way \citet{ost_pee_73} deduced that there must be a significant amount of dark mass hidden in the Galactic halo in order to keep the Milky Way stable from forming a bar, well before a dark matter halo was firmly inferred from flat rotation curves of many nearby galaxies.

%The forbidden velocity regions in the $l-v$ diagram are due to non-circular motions driven mainly by the large-scale non-axisymmetric structures such as the Galactic bar and spiral arms.
The first hint of a bar structure in the Milky Way was the substantial departures from circular motions of HI gas in the central parts of the Galaxy. The Leiden astronomers \citep{oor_rou_59, rou_oor_60} explained the observed HI gas in the ``forbidden regions'' of the longitude-velocity ($l-v$) diagram as an ``expanding arm'' or ``expanding ring'' from the center. \citet{devauc_64} correctly interpreted these observations as the non-circular gas kinematics induced by a central bar. \citet{binney_etal_91} demonstrated convincingly that many important central gas features on the $l-v$ diagram can be explained with the main orbital families in a barred galaxy. Nowadays the non-circular gas features in the $l-v$ diagram are interpreted through more sophisticated gas dynamical models, thus providing important constraints on the properties of both the bar and spiral arms (see \S\ \ref{sect:gas_kine} for more details).

\begin{figure}[h!]
\centering
\includegraphics[width=0.8\textwidth, angle=0]{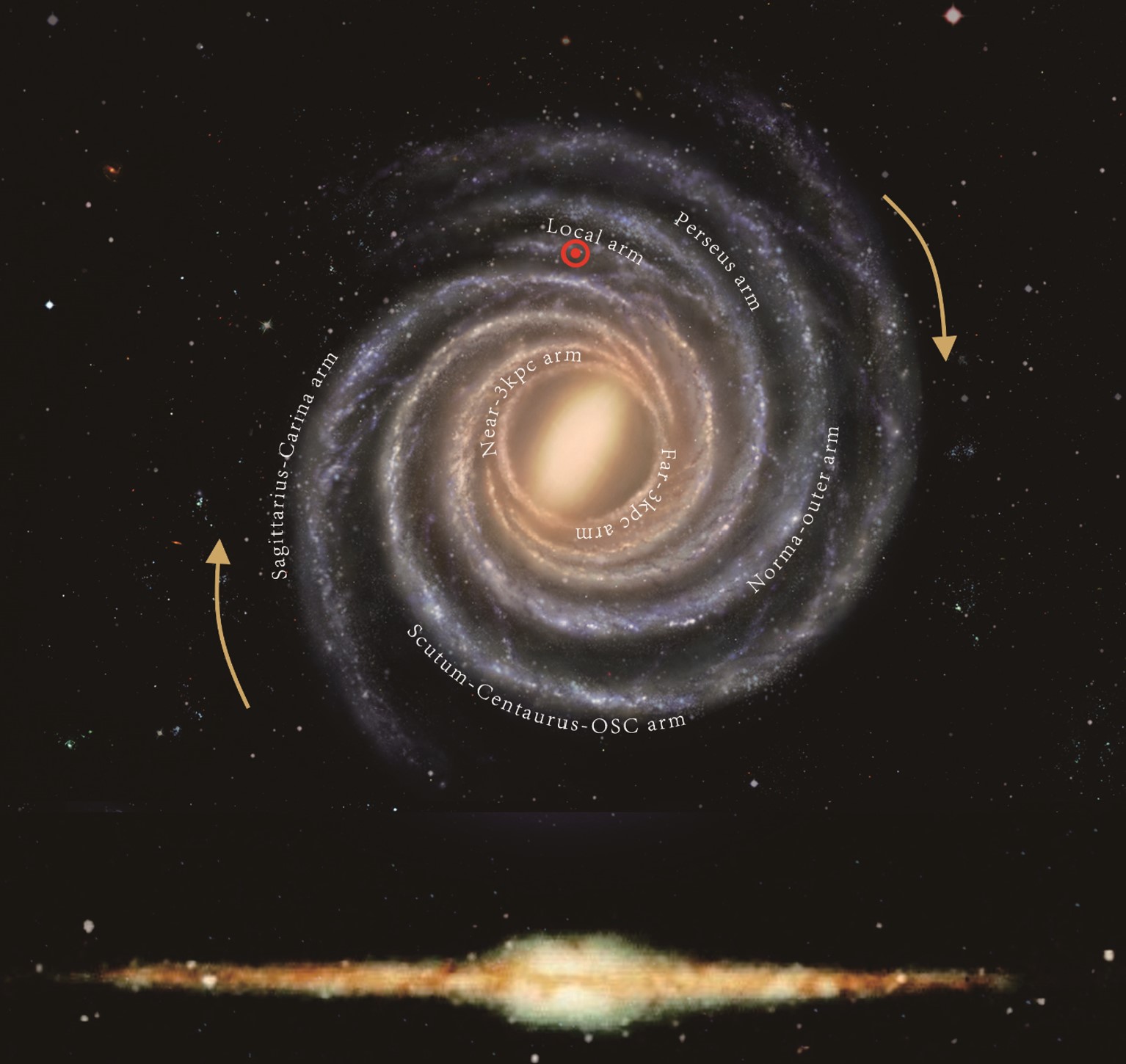}
\caption{{\it Top}: the conceptual picture of the Milky Way with its bar, four major spiral arms, a subsidiary Local arm, and 3-kpc arms. This artistic visualization also contains various important components such as gas, dust, molecular clouds and filaments, HII regions, young OB stars, and young star clusters. The Sun is marked with a circled red dot in the Local arm. The bar angle between the bar major axis and the Sun-Galactic center line is around $25 - 30^\degree$. The Galactic rotation is in the clock-wise direction. (Credit: Xing-Wu Zheng \& Mark Reid  BeSSeL/NJU/CFA). {\it Bottom}: the Milky Way seen in the infrared band by Diffuse InfraRed Background Experiment (DIRBE) on board NASA's COBE satellite (left-right flipped from the original image to be more consistent with the face-on picture).}
%https://apod.nasa.gov/apod/ap950908.html
\label{fig:complete}
\end{figure}

The near-infrared images from NASA's COBE satellite revealed clearly that the Milky Way contains an asymmetric parallelogram-shaped boxy bulge in the center \citep[][also Figure~\ref{fig:complete}]{weilan_etal_94}. The asymmetry may be explained by a tilted bar; the near end of the bar is closer to us than the far side, consequently it appears to be bigger and taller than the other side \citep{bli_spe_91}.  Although the structural parameters and orientation of the Galactic bar, mapped with various stellar tracers, are still being actively updated (see \S\ \ref{sect:3d_structure} for more details), the existence of a bar in the Milky Way has been firmly established since COBE. In \S\ \ref{sect:xshape} we review the properties of the intriguing ``X-shaped structure'' in the Galactic bar/bulge discovered about ten years ago. We discuss briefly in  \S\ \ref{sect:barmodels} the result of dynamical models of the Milky Way bar. The measurement of the Galactic bar pattern rotation speed  using stellar and gas kinematic data is reviewed in \S \ \ref{sect:pattern}.

%The chemical composition and the stellar age are crucial parameters to constrain galaxy formation history and to establish the connection to stellar populations in other structural components.
The structural components of the Galactic bulge/bar also have to be understood in the context of the chemical composition and the age of the bulge stars, which contain key information constraining the formation history of the Milky Way. Bulge stars have a broad metallicity distribution \citep{mcw_ric_94, zoccal_etal_08} and are $\alpha$-enhanced. The bulk of bulge stellar population is as old as $\sim 10$ Gyr \citep[e.g.,][]{ortola_etal_95,lecure_etal_07,clarks_etal_08,valent_etal_13,hassel_etal_20}, including some of the oldest stars in the Milky Way \citep[e.g.][]{howes_etal_14,sch_cas_14}. So most bulge stars must have experienced a rapid, early formation, yet it becomes unambiguous that they are part of the boxy bulge/bar from star counts (see \S\ \ref{sect:3d_structure}) and dynamical models. In \S\ \ref{sect:barmodels} we also review the (chemo-)dynamical models of the Galactic bulge/bar attempting to link its main dynamical and chemical properties.

%The metallicity distribution in the bulge displays a vertical gradient both on and off the minor axis away from the Galactic plane \citep{zoccal_etal_08,gonzal_etal_11a,johnso_etal_11, johnso_etal_12,johnso_etal_13}, while \cite{rich_etal_12} found no major vertical abundance gradient close to the disk plane ($b\le 4^\degree$).

%Compared to metallicity, age determination is more difficult and imprecise (see the review by \citealt{soderb_10} and references therein). The bulk of bulge stars is old ($\sim 10$ Gyr) \citep[e.g.,][]{ortola_etal_95,lecure_etal_07,clarks_etal_08}. However, intermediate-age metal-rich stars are also detected in the bulge region, with the exact relative fraction still under debate \citep{clarks_etal_11, bensby_etal_11, bensby_etal_12,ness_etal_14}. In addition, there is a nuclear disk of much younger stellar population with ongoing star formation in the central 200 pc (sometimes termed ``nuclear bulge"), whose mass is about $1.5\times 10^9 \Ms$ \citep{launha_etal_02}.
% and may include some intermediate-age metal-rich stars whose exact fraction is still under debate \citep{clarks_etal_11, bensby_etal_11, bensby_etal_12,ness_etal_14}.

Figure~\ref{fig:complete} shows an artistic impression of the majestic Milky Way structure viewed face-on, along with the infrared observation by COBE as viewed from our own solar perspective. The Milky Way probably has a strong peanut-shaped bar with two pairs of spiral arms and a subsidiary Local arm (see \S\ \ref{sect:spirals}). Its Hubble type may be somewhere between SBb and SBc types.
%The formation scenarios of spiral arms are discussed in \S\ \ref{sect:spiral_theory} before we conclude the review in \S\ \ref{sect:summary}.

\section{The Galactic Bar}
\label{sect:bar}

The study of the Galactic bar/bulge has progressed enormously in the last ten years thanks to many large surveys and sophisticated (chemo-)dynamical tools striving to model these large datasets. We summarize the main results to date on the structure and kinematics of the Galactic bar. Other extensive reviews of the Galactic bar/bulge can be found in \citet{rich_13, gon_gad_16, bla_ger_16, barbuy_etal_18}.

\subsection{Basic Structural Properties}
\label{sect:3d_structure}

%Include a table showing the bar length; axial ratio; pattern speed; bar angle.

The Milky Way contains an asymmetric box-shaped bulge \citep{maihar_etal_78,weilan_etal_94, dwek_etal_95}. The connection of this boxy structure with an edge-on bar is strong; the COBE infrared image in Figure~\ref{fig:complete} shows a parallelogram-shaped distortion that is naturally explained by a bar as a perspective effect: the near end of the bar (positive longitude $l$) is closer to us than the far end (negative longitude $l$). Consequently the vertical extent of the bulge is greater on the near side than on the far side \citep{bli_spe_91}. The case became even more compelling when \citet{shen_etal_10} built a simple $N$-body model of the Galaxy that self-consistently develops a bar. Not only their thickened bar, as seen from the Sun, resembles the boxy bulge of our Galaxy, the model also matches the BRAVA \citep{rich_etal_07} stellar kinematic data covering the whole bulge strikingly well with no need for a classical bulge made in prior mergers. Thus it is quite likely that the bulk of the bulge is simply an edge-on triaxial bar (more details in \S~\ref{sect:barmodels}).

As in external barred galaxies, the Galactic bar consists predominantly of stars instead of gas or dust. Thus the structural properties of this triaxial bar/bulge can be directly determined more precisely from observations if one can find a good standard candle to trace the bar. As of now the best tracer to study the structure of the Galactic bulge is red clump giants (RCGs). RCGs are Helium-core burning stars, and are the metal-rich equivalent of the horizontal branch stars. RCGs have a narrow range of absolute magnitudes and colors with weak metallicity dependence \citep[e.g.][]{zhao_etal_01,sal_gir_02}, thus can be used as a good distance indicator. They are also abundant enough and sufficiently bright to be seen out to the Galactic bulge, thus well-suited to investigate the structural properties of the bulge. From the magnitude distribution of the RCGs one can derive line-of-sight densities and combine many line-of-sight density measurements to get the full three-dimensional density distribution of the bulge.

Using 0.7 million RCGs from the Optical Gravitational Lensing Experiment (OGLE) project, \citet{stanek_etal_97} modelled the Galactic bar by fitting the observed luminosity functions in the red clump region of the color-magnitude diagram. Their models have a bar angle of $20^\degree - 30^\degree$ (defined as the angle between the bar major axis and the Sun-Galactic center line), with axis ratios corresponding to $x_0:y_0:z_0 = 1.00:0.43:0.29$ ($x_0, y_0$ are the semi-major, semi-minor bar axis scalelengths in the Galactic plane, respectively, and $z_0$ is the vertical bar scalelength). \citet{cao_etal_13} updated the bar structural parameters with nearly 3 million stars in OGLE III survey. They found a nearly prolate bar with an axial ratio of $x_0: y_0: z_0 \approx 1.00: 0.43: 0.40$ with $x_0= 0.67\; \kpc$. Their bar angle is $29\pm 2^\degree$, slightly larger than the value obtained from a similar study based on OGLE-II data \citep{ratten_etal_07b}.

The Vista Variables in the Via Lactea (VVV) survey \citep{saito_etal_12} offers a unique opportunity to study the Galactic bulge. The depth of VVV exceeds that of Two Micron All Sky Survey (2MASS) by $\sim 4$ mag, allowing the detection of the entire RCG population in most of the bulge region except the most highly extinct and crowded regions ($|b|<1^\degree$). With nearly 8 million RCGs from VVV, \citet{weg_ger_13} measured the three-dimensional density distribution of the Galactic bulge  covering the inner $(\pm2.2\times \pm 1.4 \times \pm 1.1) \;\kpc$.  Their measurement is non-parametric with an assumption that the three-dimensional bulge is eightfold mirror triaxially symmetric. They found a bar angle of $27 \pm 2^\degree$. The resulting density distribution shows a highly elongated bar with face-on projected axis ratios $\approx (1 : 2.1)$ at $\sim 2 \;\kpc$ along the bar major axis. The density falls off roughly in an exponential manner along the bar axes, with axis ratios $(1.00 : 0.63 : 0.26)$ and exponential scalelengths $(0.70 : 0.44 : 0.18) \;\kpc$. The axial ratio varies with radius as the true shape of the bar deviates from an ellipsoid to become peanut-shaped (see \S\ \ref{sect:xshape}).

Unlike \citet{weg_ger_13}, \citet{simion_etal_17} adopted a parametric approach to model the RCGs in VVV with an analytic function that describes the full 3D bulge density distribution summed to a background population consisting of the thin and thick disks, generated with Galaxia \citep{sharma_etal_11}.  For the bulge density distribution they tested an exponential-type model, a hyperbolic secant density distribution and a combination of the two. The best-fit parametric model of the bulge density is exponential with an axis ratio of $(1.00:0.44:0.31)$ and provides a good fit with a median percentage residual of 5\% over the fitted region. Describing the stellar distribution in the bulge with an analytic function clearly gives a more portable solution which can be straightforwardly used in other bar/bulge dynamical modeling. They found that there exists a strong degeneracy between the bar angle and the dispersion of the RCG absolute magnitude distribution. \citet{simion_etal_17} found the bar angle to be at least $20^\degree$, which is, however, strongly dependent on the assumptions made about the intrinsic luminosity function of the bulge. \citet{shen_etal_10} also provided some constraints on the bar angle. They found that their best model tends to prefer a bar angle of $\sim20^\degree - 30^\degree$ to match the velocity profiles and the photometric asymmetry. This angle agrees reasonably well with the other independent studies.

\begin{figure}[h!]
\centering
\includegraphics[width=0.6\textwidth, angle=0]{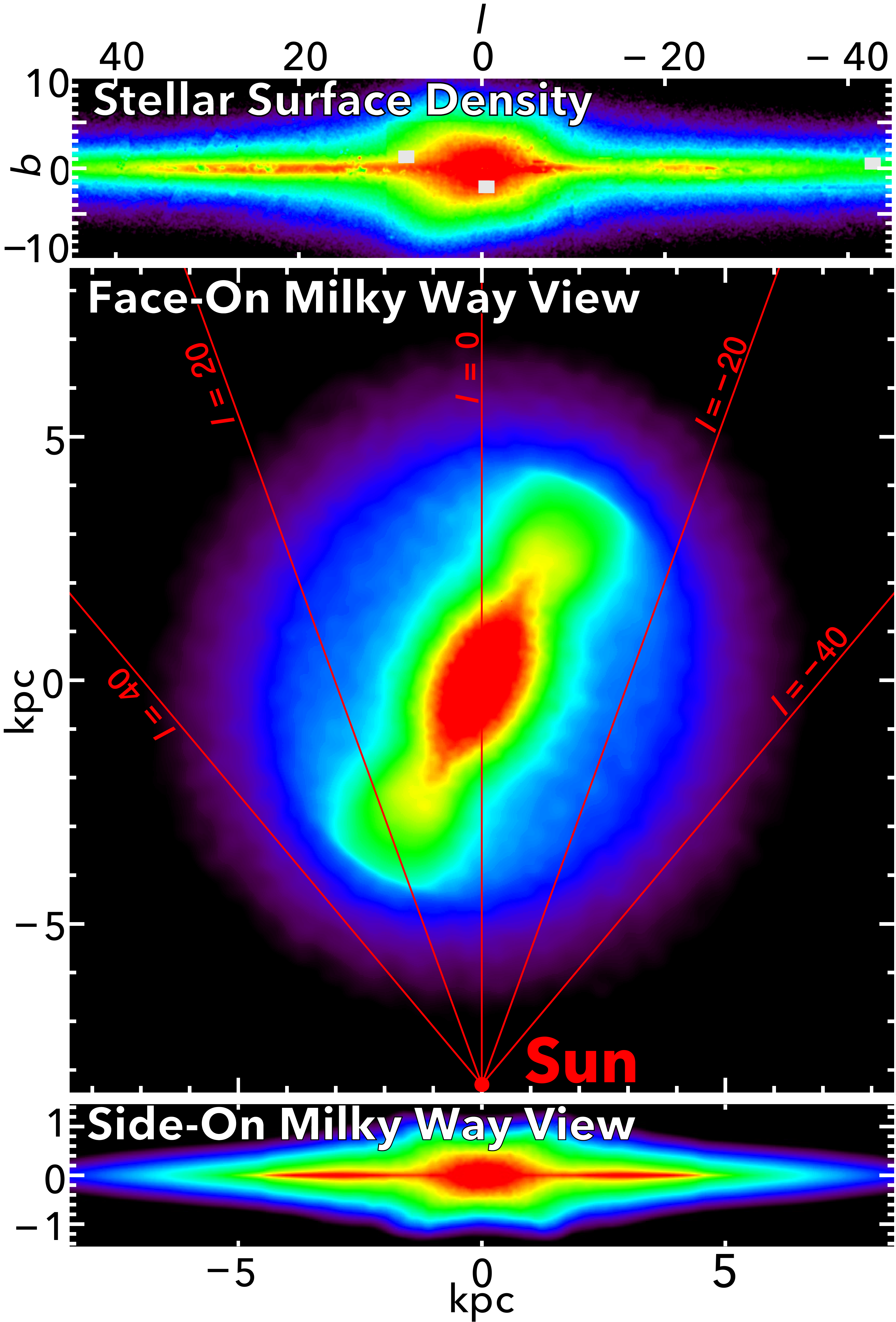}
\caption{The Galactic boxy bulge and long bar reconstructed by combining various NIR surveys. {\it Top}: the inner Galaxy in solar perspective. {\it Middle}: Face-on projection of best-fitting RCG star count model. {\it Bottom}: side-on view of the bar/bulge along the intermediate axis. Adapted from \citet{wegg_etal_15}.}
\label{fig:wegg15}
\end{figure}

Based on stars counts from the Spitzer/GLIMPSE survey, \cite{benjam_etal_05} argued for another planar long bar passing through the GC with half-length 4.4 kpc tilted by $\sim 45^\degree$ from the Sun-GC line (dubbed as the ``long bar''), which seems misaligned with the boxy bulge bar (see also \citealt{cabrer_etal_07}). If the large misalignment were real, then the co-existence of the long bar with the similarly-sized bulge bar is dynamically puzzling as their mutual torque tends to align the two bars on a short timescale, unless their size ratio is extreme ($0.1\sim 0.2$) as in some double-barred galaxies \citep{erw_spa_02,deb_she_07,she_deb_09}. \citet{wegg_etal_15} investigated in greater detail the Galactic long bar outside the bulge, using a larger and more uniform RCG combined sample from United Kingdom Infrared Deep Sky Survey (UKIDSS), 2MASS, VVV, and GLIMPSE surveys. They found that the long bar extends to $l\sim 25^\degree$ at $|b| \sim 5^\degree$ and to $l\sim 30^\degree$ at lower latitudes. The bar angle of the long bar is about  $29.5^\degree$, nearly aligned with the boxy bulge/bar at $|l| < 10^\degree$. The best model in \citet{wegg_etal_15} at various projections is shown in Figure~\ref{fig:wegg15}. The scale height of RCG stars smoothly transitions from the boxy bulge to the thinner long bar, indicating that the boxy bulge and the thin long bar may be different components of the same coherent bar structure as seen in simulations \citep[e.g.,][]{athana_05, mar_ger_11,li_she_15} and in some external galaxies. There seem two scale heights in the long bar: a 180 pc thin bar component and a 45 pc ``superthin'' bar components which exist predominantly towards the bar end. They also constructed parametric models for the red clump magnitude distributions and find a total bar half-length of $5.0 \pm 0.2 \;\kpc$ (including the super thin bar component). Thus the boxy-peanut barred shape in the inner $\sim2 \;\kpc$  transits smoothly outwards into a long thin bar with a half-length of $5.0\pm0.2 \;\kpc$, with a consistent bar angle of $\sim28^\circ$ or so.

%``These recent results are based on a larger and more uniform data base and on a more uniform analysis than the earlier work on the long bar, using cross-checked star-by-star extinction corrections and a statistical rather than CMD-based selection of RCG stars. This leads to smaller errors in the RCG magnitude distributions and reduced scatter between neighbouring fields, particularly near the Galactic plane. These results therefore supercede in particular the earlier claim that the long bar is an independent bar structure at angle 45 degrees and misaligned with the b/p bulge.''

%\Comment{\Bulge structure traced by RR Lyrae, and its rotation, possibly a classical bulge or inner halo? Kunder et al. 2016, Du Hangci paper. \citet{pietru_etal_14} }  Simion (2020) bar angle?

\subsection{Peanut-/X-shaped Structure}
\label{sect:xshape}

\citet[][hereafter MZ10]{mcw_zoc_10} and \citet{nataf_etal_10} reported a clear double-peaked magnitude distribution of the RCGs in many Galactic bulge fields (often termed the ``split red clumps'') in the 2MASS and OGLE data, respectively. This phenomenon of split red clumps was initially puzzling. Since RCGs are a good distance indicator, MZ10 suggested that the bimodality is hard to explain with a naive tilted ellipsoidal bar since the line of sight crossing the bar can only result in stars with one distance. \citet{nataf_etal_10} speculated that one RCG population belongs to the bar and the other to the spheroidal component of the bulge.  Another puzzling fact is that distances of the bright and faint RCGs are roughly constant at different latitudes, which was hard to understand with a naive straight bar. MZ10 proposed that the observed evidence can be well explained with a vertical ``X-shaped structure'' in the bulge region. The existence of this structure was later confirmed by various groups \citep{saito_etal_11,ness_etal_12,nataf_etal_15}. They found that the X-shaped structure exists at least within $|l| \leq 2^\circ$, and displays front-back symmetry. At $|b| <  5^\degree$, two RCGs begin to merge due to severe dust extinction and foreground contamination (MZ10; \citealt{weg_ger_13}). Incorporating proper motions from the VVV Infrared Astrometric Catalogue (VIRAC) and {\textit Gaia} DR2, \citet{sander_etal_19} and \citet{clarke_etal_19} verified that the differential rotation between the double peaks of the magnitude distribution of RCGs indeed confirms the X-shaped nature of the bar-bulge, thus ruled out the alternative explanation that the observed split red clumps is  due to  a population effect \citep{lee_etal_15}.

\begin{figure*}[!ht]
\centerline{\includegraphics[angle=0.,width=0.6\hsize]{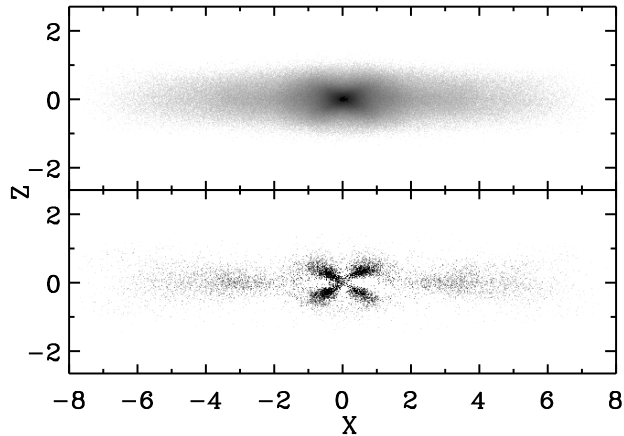}}
\caption{Demonstration of the X-shape structure in the S10 bar model. The upper panel shows the side-on view of the model and the
lower panel shows the residual after fitting and subtracting the underlying smooth light contribution. The vertical X-shaped structure is highlighted in
this residual image. The length unit is $R_{\rm d} = 1.9 \rm \; kpc$. Reproduced from \citet{li_she_12}.}
\label{fig:xshape_l}
\end{figure*}

%confirmed also in the ARGOS sample \citep{ness_etal_12}. The RCGs seem to be distributed in a vertically-extended X-shaped structure (MZ10). ; it seemed difficult to explain them using a tilted naive ellipsoidal bar.

% This X-shaped structure does not have a straight-forward explanation in classical bulge formation scenarios, but it is a natural outcome of bar thickening processes.
The X-shaped structure cannot be explained straightforwardly in classical bulge formation scenarios, but it can develop naturally in the bar thickening process. A realistic bar is not a purely triaxial ellipsoid since it usually thickens through the vertical buckling instability or resonant trapping after a cold disk suffers from the in-plane bar-forming instability. As a result, a steady bar often acquires a boxy/peanut shape after the dynamical instabilities. This is relatively well known in both the bar dynamics community  \citep{com_san_81, combes_etal_90, raha_etal_91,pfe_fri_91,athana_05,martin_etal_06,shen_etal_10} and the external galaxies with a boxy/peanut-shaped bulge \citep[e.g.,][]{bur_fre_99, bureau_etal_06, laurik_etal_14}.

The Milky Way bar is no exception to the peanut shape. A buckled bar in numerical simulations naturally reproduces the observed X-shape properties in many aspects \citep{li_she_12, ness_etal_12}. \citet{li_she_12} analysed the best-fitting Milky Way bar/bulge model in S10 and found that an X-shaped structure is clearly recognizable in the  side-on view (top panel of Figure~\ref{fig:xshape_l}). They also demonstrated that it can qualitatively reproduce many observational results of the X-shaped structure, such as the double-peaked distribution in distance histograms (MZ10, \citealt{nataf_etal_10}) and number density maps \citep{saito_etal_11}. The bottom panel of Figure~\ref{fig:xshape_l} highlights the nearly symmetric ``X-shaped structure'' after the underlying smooth component is subtracted from the side-on bar model. The extent of the ``X-shape'' is roughly 3 kpc and 1.8 kpc along the bar major axis and in the vertical direction, respectively. The X-shape has a similar tilting angle as the bar, but extends to only about half the bar length.   As in observations, at a given latitude $b$ the peak positions are roughly constant, and the further peak becomes gradually stronger at decreasing $l$; at a given longitude $l$ the separation of the two peaks increases as $|b|$ increases.  \citet{li_she_12} estimated that the light fraction of this X-shaped structure is about 7\% of the whole bulge.  \cite{portai_etal_15a} performed more sophisticated modeling based on the reconstructed bulge volume density from VVV survey \citep{weg_ger_13},  and found an off-centered X-shape comprising about 20\% of the bulge mass.

\begin{figure*}[!ht]
\centerline{ \includegraphics[angle=0.,width=\hsize]{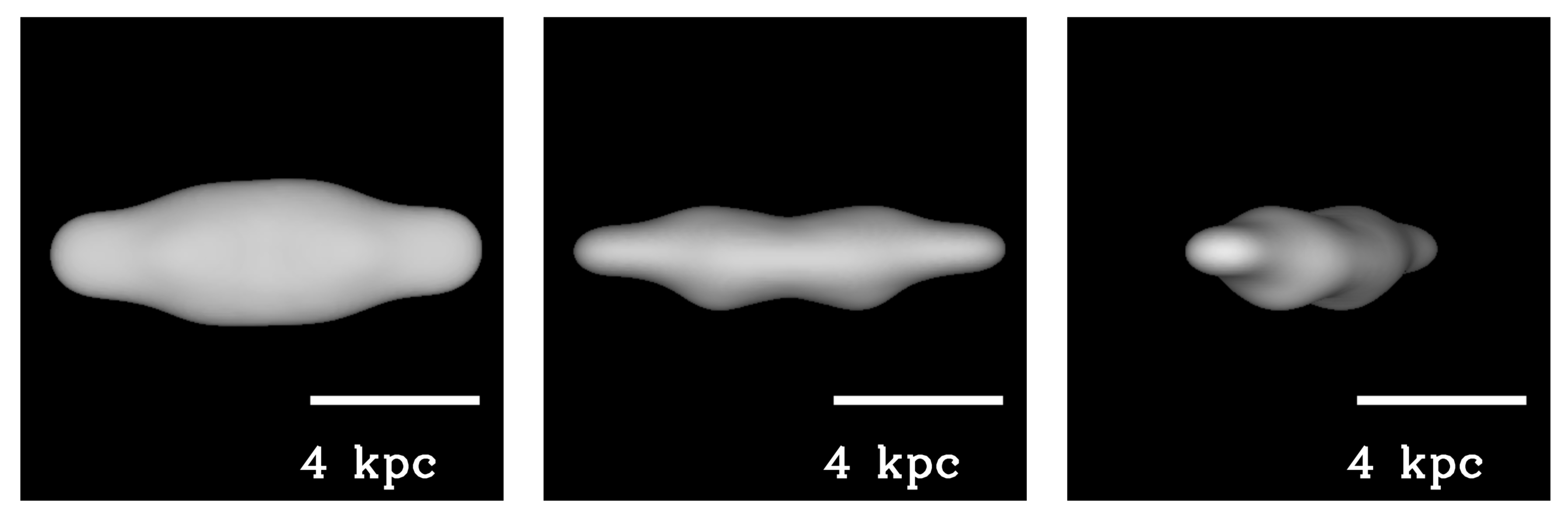}}
\caption{3D isodensity surfaces of a strongly buckled bar in $N$-body simulations. The left and middle panels show the face-on and side-on appearance of the bar, respectively. The right panel shows an edge-on view at a bar angle of $25^\degree$. Adapted from \citet{li_she_15}.}
\label{fig:3dx}
\end{figure*}

However, the true three-dimensional shape of the X-shaped structure is not as simple as a letter ``X'' with four or eight conspicuous arms. Figure~\ref{fig:xshape_l} may give such a biased impression because human eyes tend to be drawn by small-scale density enhancements. The true three-dimensional shape of the X-shaped structure is actually more like a peanut. This is demonstrated in \citet{li_she_15} who estimated the 3-D volume density of $N$-body bar models with an adaptive kernel smoothing technique \citep{silver_86, she_sel_04}. Figure~\ref{fig:3dx} shows clearly that the morphology of a strongly buckled  bar  transitions  gradually  from  a  central  boxy core  to  a  peanut  bulge,  and  then  to  an extended, thin bar.  This was also found in the observations with larger and more uniform samples of RCGs \citep{weg_ger_13, wegg_etal_15,simion_etal_17}.  But the peanut-shaped bulge can still  reproduce qualitatively the observed double-peaked distance distributions that were used to infer for the discovery of the X-shape. 
%Our visual perception of an ``X" is enhanced by the pinched inner concave isodensity contours of the inner peanut structure \citep{li_she_15}. 
\citet{li_she_15} demonstrated that the pinched concave isodensity contours of the inner peanut structure can enhance our visual perception of an letter ``X''.
Note that the central  boxy  core  is  shaped  like  an  oblong  tablet,  extending  within $\sim$ 500 pc or $|b| \sim 4^\degree$ near the Galactic plane. From the solar perspective, lines of sight passing through the central boxy core do not show bimodal distributions, in agreement with observations \citep[MZ10;][]{weg_ger_13}.

%Orbital structure studies are essential for understanding the properties of the X-shaped structure, which is the outcome of the collective buckling instability.

Until quite recently it was widely believed that the peanut-shape of a bar is supported by orbits trapped around the 3D $x_1$ family, also known as banana orbits due to their banana shape when viewed side-on \citep[e.g.,][]{patsis_etal_02,skokos_etal_02a, athana_16_springer}. The backbone orbits of a 3D buckled bar are the $x_1$ tree, i.e., the $x_1$ family plus a tree of 3D families bifurcating from it \citep{pfe_fri_91}. 
%However, recently \citet{portai_etal_15b} classified orbital families in the peanut/X-shaped bulges, and proposed that `brezel-like' orbits, which may be closely related to the $x_{1}mul_{2}$ family \citep{pat_kat_14a}, as the main contributor to the peanut shape. \citet{qin_etal_15} also found that stars in the boxy peanut-shaped bulge do not necessarily stream along simple banana orbits. 
\citet{portai_etal_15b}, however, proposed that `brezel-like' orbits are instead the main contributor to the peanut shape. Such orbits may be related to the so-called ``$x_{1}mul_{2}$'' orbit family \citep{pat_kat_14a}. \citet{qin_etal_15} also showed from the kinematics of a simulated boxy peanut-shaped bar that stars in the bar do not show a clear sign of streaming along banana orbits.
\citet{abbott_etal_17} found that only `fish/pretzel' orbits and `brezel' orbits, comprising 7.5 per cent of the total mass,  show a distinct X-shape in unsharp masked images, but nearly all bar orbit families contribute some mass to a 3D boxy peanut-shaped bar \citep[also see][]{parul_etal_20}. 
Clearly we need more in-depth investigations on the orbital structure and vertical resonant heating process \citep[e.g.][]{quille_etal_14,sel_ger_20} to make more specific predictions for the Galactic bar.
%More detailed studies on the orbital structure and vertical resonant heating \citep[e.g.][]{quille_etal_14,sel_ger_20} are clearly desired to make more specific predictions for the Milky Way.

Although the buckling instability has been demonstrated to be sufficient to thicken the Galactic bar into the peanut shape as shown in many previous studies \citep[e.g.,][]{shen_etal_10}, it is unclear if it is the only way. \citet{sel_ger_20} carefully studied three mechanisms for bar thickening: the well-known buckling instability, vertical excitation of bar orbits as a 2:1 vertical resonance sweeps out along the bar \citep{quille_etal_14},  and gradually trapped bar orbits into the 2:1 vertical resonance. They found the fourth-order Gauss-Hermite  coefficient $h_4$ profile of the vertical velocity distribution (see also \citealt{debatt_etal_05}) may be a good diagnostic to discriminate between a bar made via the buckling instability from the other two mechanisms. It remains hopeful that better proper motion data in the future will distinguish the thickening mechanism responsible for the Galactic bar. If the Galactic bar indeed experienced a buckling instability, then the X-shaped or peanut-shaped structure becomes nearly symmetric with respect to the disk plane $\sim$ 2 Gyr after the instability gradually saturates.  Thus the observed symmetry (MZ10) might imply that the X-shaped structure in the Galactic bulge has been in existence for at least a few billion years \citep{li_she_12}.

The existence of the X-shaped structure in our Milky Way provides additional evidence that the Galactic bulge is shaped mainly by internal disk dynamical instabilities instead of mergers, because no other known physical processes can naturally develop such a structure.
%\citet{deprop_etal_11} studied the radial velocity and abundances of bright and faint RCGs at $(l, b) = (0^\degree,-8^\degree)$, and found no significant dynamical or chemical differences. This may suggest that the two RCGs indeed belong to the same coherent dynamical structure, which can be naturally made in the formation of the bar/boxy bulge.

\subsection{Dynamical Models of the Galactic Bar/bulge}
\label{sect:barmodels}

%See other reviews \citet{gerhard_18_IAUS}
%$\bullet$ Early models like the Zhao 1996 model

Building a dynamical model of the Milky Way bulge/bar requires stellar or gas kinematics as constraints. Based on the Schwarzschild orbit-superposition method, \citet{zhao_96} developed the first 3D rotating bar model that fits the density profile of the COBE light distribution and scarce kinematic data at Baade's window. His model then was constructed with 485 orbit building blocks, and little stellar kinematic data were available to explore the uniqueness of this model, unfortunately.

%\citet{hafner_etal_00} extended the classical Schwarzschild technique by combining a distribution function that depends only on classical integrals with orbits that respect non-classical integrals, i.e., Schwarzschild's orbits were used only to represent the difference between the true galaxy distribution function and an approximating classical distribution function. They used the new method to construct a dynamical model of the inner Galaxy with an orbit library that contains about 22,000 regular orbits. For definiteness, they assumed a bar angle of $20^\degree$ and bar pattern speed of $60\;\kmskpc$. The model reproduced the 3D mass density obtained through deprojection of the COBE surface photometry, and the then-available kinematics within the bar corotation radius (3.6~kpc).

%$\bullet$ Cylindrical rotation and BRAVA

The Bulge Radial Velocity Assay (BRAVA) project aims to study the stellar kinematics covering the whole Galactic bulge with $\sim9000$ M giants as tracers \citep{rich_etal_07}. These giants provide most of the box-shaped 2 $\mu$m light distribution that hints for the presence of a bar. The BRAVA results show that the boxy bulge rotates nearly cylindrically, i.e., rotation is roughly constant regardless of the height above the disk plane \citep{howard_etal_09, kunder_etal_12}. BRAVA kinematics also put the Galactic bulge near the ``oblate isotropic rotator'' line in the so-called Binney's $V_{\rm max}/\sigma-\epsilon$ diagram \citep{binney_78}, which shows that the bulge is not a hot, slowly-rotating system supported by random motions.  The Abundance and Radial velocity Galactic Origins Survey (ARGOS) obtained radial velocities and stellar parameters for an even larger sample of 28000 stars in the bulge and inner Milky Way \citep{freema_etal_13}. The clear cylindrical rotation of the bulge was confirmed in the ARGOS  \citep{ness_etal_13a} and the Giraffe Inner Bulge Survey (GIBS) data \citep{zoccal_etal_14}. These much better kinematic data with more complete spatial coverage provide crucial constraints for better dynamical models.

%%% $\bullet$ N-bodys Shen model, buckling and its success, other $N$-body, Di Matteo  etal. Debatissta

\begin{figure}[ht!]
\centerline{\includegraphics[width=.8\hsize]{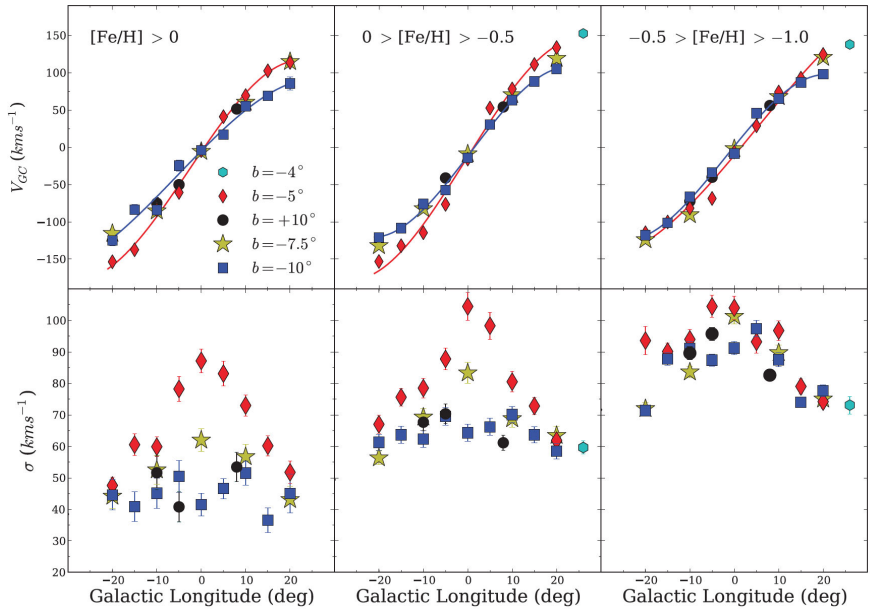}}
\caption{Rotation and velocity dispersion profiles in the ARGOS observations towards the Galactic bulge fields from \cite{ness_etal_13a}. The three columns correspond to three different metallicity bins, decreasing from left to right. Different symbols represent stars in different fields. Reproduced from \citet{ness_etal_13a}.}
%\vskip -15pt}
\label{fig:ness_kine}
\end{figure}

$N$-body simulations of the Galactic bar/bulge have provided insight on its formation and evolution.  For example, the S10 $N$-body model was initially designed to match the BRAVA data without too many free parameters to tweak. It is one of the simplest evolutionary bar models that developed naturally from the bar instability of a cold massive precursor disk. Despite its simplicity, it has enjoyed successes in many aspects. The physical processes shaping the structural formation of the Galactic bar/bulge are well understood; the in-plane bar instability gives rise to a massive bar that then got thickened vertically into a boxy/peanut/X shape in the subsequent firehose/buckling instability (see \citealt{sellwood_14} for a comprehensive review on these instabilities). The best-fitting model of S10 also naturally reproduces many other observational results reasonably well, e. g., excellent match to the stellar kinematics, the bar angle of $20^\degree-30^\degree$, a reasonable bar length ($\sim 4 \;\kpc$), a bar pattern speed of $\sim 40 \;\kmskpc$, the vertical metallicity gradient (\citealt{mar_ger_13}, \citealt{liu_she_21}), and gives an upper mass limit on a possible classical bulge. A drawback of S10 model is its adoption of a rigid dark matter halo for simplicity, thus it omitted the dynamical friction between the bar and halo which may affect the long-term evolution of the bar. Fortunately, the low mass fraction of a cored dark matter halo in the bar region \citep{portai_etal_17a} warrants that such bar-halo dynamical friction may not be too strong in the Milky Way. Simple bar models like S10 may serve as a physically-motivated starting point, then more chemo-dynamical complexities of the Milky Way bulge may be gradually incorporated into it. \citet{dimatt_etal_15}, \citet{fragko_etal_18}, and \citet{dimatt_etal_19} showed further how adding a second, thick disk in their $N$-body simulations may further improve the chemo-kinematic relations we describe below (Figure~\ref{fig:ness_kine}). \citet{debatt_etal_17} demonstrated how initially co-spatial stellar populations with different in-plane random motions separate when a bar forms. Although $N$-body simulations can provide the full evolutionary history from plausible initial conditions, they are inflexible in the sense that numerous trials of different initial configurations are required to reproduce the desired results, which are not always controllable, thus limiting the systematic exploration of parameter space to match the observational results.

% Di Matteo 2016 PASA: N-body models of a boxy-/peanut-shaped bulge formed from a thin disk through the intermediary of a bar have been successful in interpreting a number of global properties of the Galactic bulge, but they fail in reproducing the detailed chemo-kinematic relations satisfied by its components and their morphology. It is only by adding the thick disk to the picture that we can understand the nature of the Galactic bulge.

%Theoretical modeling of the Milky Way bulge made intense use of $N$-body simulations. The basis of a successful Galactic bulge model is a fully evolutionary bar model that developed naturally from the bar instability of a cold massive disk. A successful high-resolution bar model was developed in \citet[hereafter S10]{shen_etal_10}, which was initially motivated to match the BRAVA stellar kinematic data. $N$-body bar models to explain the Galactic bulge were already attempted in early studies such as \cite{fux_97,sevens_etal_99}, but little stellar kinematic data were available to constrain their models.

%%% $\bullet$ Portail M2M models.

The made-to-measure (M2M) method \citep{sye_tre_96} is a complementary alternative to $N$-body models, and is more flexible in steering models to match a large number of data constraints. In this approach one first constructs a reasonable $N$-body model with the essential physics to match the galaxy under study.  The weights of the particles are slowly adjusted as particles proceed in their orbits until the time-averaged density field and other observables converge to the observational value, through a weight evolution equation according to the mismatch between the model and target observables.  The M2M method has been continuously tested and improved in various implementations \citep[e.g.][]{bissan_etal_04,delore_etal_07,dehnen_09,lon_mao_10,lon_mao_12,hun_kaw_13,portai_etal_15a, lon_mao_18}, and has become an important tool in the dynamical modelling of the Milky Way bar/bulge. \citet{portai_etal_17a} built made-to-measure dynamical models that fit the RCG density from the VVV, UKIDSS, and 2MASS survey, kinematics from BRAVA, ARGOS, and OGLE surveys. Their models gave a bar pattern speed of $39.0\pm 3.5 \;\kmskpc$. The total dynamical mass in their model for the bulge volume ($\pm2.2 \times \pm 1.4 \times \pm1.2$ kpc along the bar principal axes) is $1.85\pm0.05\times10^{10}\Ms$, with a low dark matter fraction of $17\pm2\%$. Their results also implied a core or shallow cusp profile of the dark matter halo inside $\sim2\;\kpc$. The stellar mass inside the bulge volume is  $\sim 1.52 \times10^{10}\Ms $, roughly consistent with stellar mass estimate ($\sim 2.0 \pm 0.3 \times10^{10}\Ms$) by \citet{valent_etal_16} within $|b| < 9.5^\degree$ and $|l| < 10^\degree$ considering that it is a bigger spatial volume than in \citet{portai_etal_17a}.
%%% It is also worth noting that microlensing time-scale distribution prefer a Kroupa IMF in the bulge region as reported in \citet{wegg_etal_17}.

% Ness figure

%$\bullet$ chemo-kinematic properties, Ness figure. Bar may not be the whole bulge
Observations also reveal distinctly different kinematic properties between the relatively more metal-rich and metal-poor stars in the Galactic bar/bulge. For example, \citet{babusi_etal_10} showed that the more metal-rich population has bar-like kinematics and the more metal-poor population is likely associated with an old spheroid or a thick disk \citep[also see][]{hill_etal_11, rojas_etal_14}. The metal-rich population demonstrates smaller velocity dispersion and is less $\alpha$-enhanced compared to the metal-poor one \citep{johnso_etal_11, uttent_etal_12, zoccal_etal_17}, although the rotation curves are similar for three different metallicity bins (Figure~\ref{fig:ness_kine}). Based on the ARGOS sample, \cite{ness_etal_12} found that the X-shaped structure is shown in the metal-rich stars ($\rm [Fe/H]>-0.5$) rather than the metal-poor ones ($\rm -1.0<[Fe/H]<-0.5$) (also see \citealt{uttent_etal_12,gilmor_etal_12}), indicating the more metal-rich stars are pre-dominantly tracing the boxy bar/bulge.
%\cite{ness_etal_13a} found that stars with $\rm [Fe/H] > -0.5$ are part of the boxy bar/bulge, and the metal-poor stars are likely associated with thick disk.
%\citet{ness_etal_13a} also found  stars with $\rm [Fe/H] \sim +0.15$ are more prominent close to the plane than the metal-poor stars, appearing as a vertical abundance gradient of the bulge.
% Across the bulge fields in ARGOS survey, the metal-poor population ($\rm [Fe/H] < -1.0$) was also found to be kinematically distinct with large velocity dispersion and non-cylindrical rotation \citep{ness_etal_13b}.

These chemo-kinematic relations set key constraints for dynamical models including chemical information. \citet{portai_etal_17b} built a self-consistent chemodynamical model to fit observational results for the galactic bulge, bar, and inner disk. They extended the M2M dynamical model from \citet{portai_etal_17a} to reproduce the observed metallicity-dependent density distribution and kinematics. They found most metal-rich stars ($\rm [Fe/H] \ge -0.5$) belong to the bar component, while the metal poor stars ($\rm [Fe/H] \le -0.5$) outside the central kpc are more likely to have a thick disk origin. Their model could also reproduce the observed vertex deviations in Baade's window \citep{soto_etal_07,babusi_etal_10}. As proper motions from {\textit Gaia} and VIRAC and other chemical abundances from large spectroscopic surveys are being included as model constraints, chemo-dynamical models will become more powerful in revealing the detailed dynamics and formation history of the Milky Way bulge/bar.

\subsubsection{Is There a Classical Bulge Component?}
\label{sect:classical}

%What is a classical bulge?
%Old? Spheroidal shape? Non-rotating?
%Classical bulges ≈ Mini-ellipticals
%“fundamental plane” of ellipticals: a tight relation of (size, 𝜎, surface brightness)
%No freedom to tinker classical bulge profiles to make them easy to hide.
%A smaller CB  higher central surface brightness, much denser than a star cluster (4.2pc) or disky pseudo- bulge (~45pc)

%Galactic bulges contain crucial information about the galaxy formation and evolution. Major mergers between galaxies generally create a significant classical bulge that is similar to ellipticals in many aspects, whereas the long term internal secular evolution in the disk galaxy tends to build up a pseudobulge \citep{kor_ken_04}.

%Galactic bulges contain crucial information about the galaxy formation and evolution. Major mergers between galaxies generally create a significant classical bulge that is similar to ellipticals in many aspects, whereas the long term internal secular evolution in the disk galaxy tends to build up a pseudobulge \citep{kor_ken_04}.

Key information about the galaxy formation and evolution may be learned by studying galactic bulges. It is generally expected that major mergers between galaxies tend to create a classical bulge bearing a resemblance to ellipticals in many aspects, while the slower internal secular evolution processes in a disk galaxy \citep{sellwood_14} tend to build up a disk-like ``pseudobulge'' \citep{kor_ken_04}.

Now it is widely accepted that the bulk of the Galactic bulge is actually the boxy/peanut-shaped part of the Galactic bar. Although most bulge stars are old ($>$ 10 Gyr), but the assembly epoch of the bulge structure itself may not be that ancient. There are a number of reasons why an old classical bulge made in prior mergers or a monolithic collapse is unlikely to be significant in the Milky Way.

First, a classical bulges is like a mini-elliptical galaxy, which stays on the ``fundamental plane'' of ellipticals, i.e., the tight correlations between galaxy size, central surface brightness, and velocity dispersion. So there is not much freedom to postulate classical bulges having arbitrary properties, such as low surface brightness, to make them hard to detect \citep{kormen_etal_10}. A very low-mass classical bulge tends to have high surface brightness and large effective radius predicted from fundamental plane correlations, and it is much denser than a star cluster or a disky pseudo-bulge. \citet{kor_ben_19} estimated that a classical bulge with a bulge-to-total luminosity ratio $B/T=0.02$  corresponds to an absolute magnitude of $M_V \approx -16.3$,  i.e. similar to that of M32 ($M_V \approx -16.7$), and has an effective radius of $100\sim200$ pc. The fact that we can even detect the tiny nuclear star cluster ($R_e \sim 4$~pc) in the Galactic center shows that the Milky Way does not contain a small classical bulge with a mass of $\sim2\%M_{\rm disk}$, otherwise we would have only observed a dense classical bulge instead of the nuclear star cluster.

Secondly, the metal-poor stars ($\rm -1.0<[Fe/H]<-0.5$)  in the bulge are unlikely to belong to a classical bulge. \citet{shen_etal_10} also tested whether or not a significant classical bulge is present, since it could have been spun up by the later formation of a bar, flattened thereby and made hard to detect. They found that including a classical bulge with $\ge$ 10\% of the disk mass considerably worsens the fit of the model to the data, even if the disk properties are accordingly re-adjusted. If the pre-existing classical bulge is overly massive, then it becomes increasingly hard to match both the mean velocity and velocity dispersions simultaneously \citep[see also][]{saha_etal_12}. Such a small bulge can neither explain the large fraction of metal-poor stars \citep{ness_etal_13a,zoccal_etal_18}, nor the increasing fraction with latitude $b$ \citep{dimatt_etal_14}. Also if the metal-poor stars were to be associated with a classical bulge, then the rotation velocity of the metal-poor stars will be significantly slower than that of the metal-rich ones, which is inconsistent with observations (e.g., Figure~\ref{fig:ness_kine}). In addition, a 10\%$M_{\rm disk}$  classical bulge will have a steeply rising surface brightness per the fundamental plane correlations, which is inconsistent with the nearly exponential minor-axis profile shown in the NIR data \citep{launha_etal_02}.

Thirdly, the more centrally concentrated RR Lyrae population \citep{dekany_etal_13,pietru_etal_15,kunder_etal_16,kunder_etal_20} are unlikely to be associated with the classical bulge. They only account for less than 1\% of the total mass of the bar traced by RCGs, and their metallicity (median $\rm [Fe/H] \sim -1.0$) is much more metal-poor than the bulk of the metal-poor population in GIBS and {\textit Gaia}-ESO \citep{rojas_etal_14,zoccal_etal_18}. It has been suggested that these RR Lyrae stars might  represent the inner extension of the halo confined to the inner Galaxy \citep{perezv_etal_17,savino_etal_20}, but \citet{du_etal_20} demonstrated that this is unlikely the case with a large RR Lyrae sample from OGLE-IV with {\it Gaia} DR2 proper motions.

Fourthly, \citet{clarke_etal_19} found that the proper motion correlation map displays a clear quadrupole pattern in all magnitude slices of RCGs, showing no evidence for a separate, more axisymmetric inner bulge component.

In summary, the Galactic bulge is predominantly a peanut-shaped bar that formed spontaneously from a disk, and there is no sign that the Galaxy contains a significant merger-made, classical bulge. So, from a galaxy formation point of view, the Milky Way is a ``pure-disk'' galaxy.
%
%The best-fitting model in S10 contains no classical bulge component. S10 also tested whether or not a significant classical bulge is present, since it could have been spun up by the later formation of a bar, flattened thereby and made hard to detect. They found that including a classical bulge with
%$>\sim$ 15\% of the disk mass considerably worsens the fit of the model to the data, even if the disk properties are accordingly re-adjusted. If the pre-existing classical bulge is overly massive, then it becomes increasingly hard to match both the mean velocity and velocity dispersions simultaneously \citep[see also][]{saha_etal_12}.
%
%The BRAVA kinematic observations show no sign that the Galaxy contains a significant merger-made, “classical” bulge. S10 demonstrated that the boxy pseudobulge is not a separate component of the Galaxy but rather is an edge-on bar. This result also has important implications for galaxy formation. From a galaxy formation point of view, we live in a pure-disk galaxy. Our Galaxy is not unusual. In fact, giant, pure-disk galaxies are common in environments like our
%own that are far from rich clusters of galaxies \citep{kormen_etal_10,laurik_etal_14}. Classical bulgeless, pure-disk galaxies still present an acute challenge to the current picture of galaxy formation in a universe dominated by cold dark matter; growing a giant galaxy via hierarchical clustering involves so many mergers that it seems almost impossible to avoid forming a substantial classical bulge \citep{pee_nus_10}.

\subsection{Pattern Speed of the Milky Way Bar}
\label{sect:pattern}

Bar pattern rotation speed is one of the most important parameters of the bar dynamics, as it determines the orbital structure of stars. It can be measured independently from stellar kinematics and gas kinematics, respectively.

\subsubsection{Measurement with Stellar Kinematics}
\label{sect:stellar_kine}

\citet{debatt_etal_02}  applied a modified version of the Tremaine-Weinberg continuity formalism \citep{tre_wei_84} for use with line-of-sight velocities to a small sample of OH/IR stars in the inner Galaxy. They obtained the first direct measurement of the Galactic bar pattern speed of $\omegap = (59 \pm 5) \;\kmskpc$. {\it Gaia} now provides some of the first absolute proper motions within the bulge, and the near-infrared VVV multi-epoch catalogue can complement {\textit Gaia} in highly extincted low-latitude regions. \citet{sander_etal_19} and \citet{clarke_etal_19} analysed the kinematics of the Galactic bar/bulge using proper motions from the VVV Infrared Astrometric Catalogue (VIRAC) and {\textit Gaia} DR2. The latest data with proper motions have enabled more accurate measurement of the bar pattern speed than before.

%The quantity actually measured is the difference between the pattern rotation velocity and the circular velocity at the local standard of rest (LSR).
% Debattista et al. (2002b) adapted the formalism for use with line-of-sight velocities in the Milky Way with Debattista et al. (2002b) applying the formulae to OH/IR stars across the  Galactic disk. Whilst Tremaine & Weinberg (1984) and Kuijken & Tremaine (1991) worked in 2D, we shall follow Debattista et al. (2002b) who provided expressions for 3D density distributions (but only in the case of line-of-sight velocities).

\citet{sander_etal_19_pattern} extended the revised Tremaine-Weinberg continuity formalism for use with proper motions, and derived the pattern speed of the
Milky Way’s bar/bulge.
%The method has minimal assumptions but requires complete coverage of the non-axisymmetric component in two of the three Galactic coordinates.
They  measured $\omegap = (41 \pm 3) \;\kmskpc$, which puts the corotation radius of the Galactic bar at $(5.7 \pm 0.4) \;\kpc$. They experimented the addition of data on the near or far side of the bar, and suggested a systematic uncertainty of $5-10 \;\kmskpc$ in their measurement.  \citet{clarke_etal_19} compared their stellar kinematical data to the made-to-measure barred dynamical models in \citet{portai_etal_17a}, and found that a model of the barred bulge with a pattern speed of $37.5 \;\kmskpc$ is able to match most of the observed features. These values also agree nicely with the bar pattern speed of $\sim 39-40 \;\kmskpc$ for the S10 $N$-body bar model \citep{shen_14} designed to matched the full BRAVA kinematics, and the made-to-measure models based on the mock data created from the same $N$-body model \citep{long_etal_13}.

%Including the new PM data into these made-to-measure models may give even tighter constraints on the bar pattern speed measurement.

\subsubsection{Measurement with Gas Kinematics}
\label{sect:gas_kine}

Non-circular gas kinematics was one of the first hints for the existence of a Galactic bar \citep[e.g.,][]{devauc_64}. In fact, the features in the asymmetric gas flow pattern may be used to infer the properties of the Galactic bar, especially its  pattern rotation speed. As distances to individual gas clouds are difficult to measure accurately, motions and distribution of atomic and molecular gas  are conventionally presented in the Galactic $l-v$ diagram of HI or CO gas, i.e. the plot showing how gas emission line intensity distributes in the Galactic longitude ($l$) and line-of-sight velocity ($v$) space \citep[e.g.,][]{bur_lis_93, dame_etal_01}. Most features in the $l-v$ diagram representing the dense gas distribution are driven mainly by the large-scale non-axisymmetric structures such as the Galactic bar and spiral arms. Thus $l-v$ diagram must be interpreted through careful gas dynamical models including the bar and spiral arms due to the large distance uncertainty of gas clouds, and it in turn can provide important constraints on the properties of the bar and spiral arms by matching up the simulated gas features with the observed ones.

Many hydrodynamic models of the gas flow have followed the above approach to infer the pattern rotation speed of the Galactic bar once the Galactic bar potential is constrained by star counts. The early models were able to reproduce some of the prominent features in the $l-v$ diagram, but tend to give a relatively high bar pattern speed in the range of $\omegap = 50 - 60 \;\kmskpc$ \citep[e.g.,][]{fux_99, eng_ger_99, bissan_etal_03,rod_com_08,baba_etal_10}. However, the bar pattern speed derived from gas dynamics may have some degeneracies with the properties of the bar potential such as the bar size; a more slowly-rotating longer bar might also match to the gas features in the $l-v$ diagram well. \citet{lizhi_etal_16} modelled the Milky Way gas flow pattern with a basic bar potential constrained by the density of bulge RCGs \citep{portai_etal_15a,weg_ger_13}. They found that a lower bar pattern speed may provide an even better match to the gas $l-v$ diagram than previous high pattern speed hydrodynamical simulations, reproducing features like the shape and kinematics of the Central Molecular Zone, Bania’s clumps, the connecting arm, the Near and Far 3 kpc arms, the Molecular Ring, and the spiral arm tangent points. \citet{sorman_etal_15c} also found a lower pattern speed of $\omegap = 40 \;\kmskpc$ after experimenting a range of parameters of a  rigidly rotating bar potential with only a monopole and a quadrupole components.

A drawback of the $l-v$ diagram is that it contains only the information of line-of-sight velocities, and does not constrain the tangential motions of gas. This could be further improved  by considering the latest BeSSeL results \citep{reid_etal_19}, which measures the 3D velocity and position of nearly 200 high-mass star-forming regions (HMSFRs) with high-precision VLBI data (\S\ \ref{sect:spirals}). The peculiar motions of the HMSFRs are generally small ($\sim 10\;\kms$) except in two regions; the first  is a segment of the Perseus arm that is probably in a disrupting phase \citep{baba_etal_18} and the other region is  around the bar end. The large-scale dynamics of the Galactic bar and spiral arms might be the origin for these peculiar motions. A successful dynamical model should not only reproduce the main features in the $l-v$ diagram, but also explain the peculiar motions of these observed HMSFRs. The preliminary result in \citet{lizhi_etal_21} seems to prefer a bar pattern rotation speed of $\sim 37 - 40\;\kmskpc$, and a pattern speed of spiral arms of $\sim 23\;\kmskpc$ which is less constrained by this gas model.

In summary, the most recent independent measurements of the Galactic bar pattern speed using stellar and gas kinematics appear to converge to $\sim 35 - 40 \; \kmskpc$. This $\omegap$ value corresponds to a corotation radius $R_{\rm CR} \sim 6 \;\kpc$. With a bar half-length of $R_{\rm b}\sim 5 \;\kpc$ \citep{wegg_etal_15},  we have the dimensionless $\mathcal{R} \equiv R_{\rm CR}/R_{\rm b} \sim 1.2$, which would put the Milky Way bar into the conventional ``fast bar'' ($1.0\le\mathcal{R} \le 1.4$) category \citep{deb_sel_02}.

\section{The Spiral Structure of the Milky Way}
\label{sect:spirals}

%\Comment{Intro of BeSSeL Survey.}
The BeSSeL Survey is a National Radio Astronomy Observatory (NRAO) key science project. Its aim is to determine accurate distances of HMSFRs in the bar and spiral arms of the Milky Way by measuring the trigonometric parallaxes of the methanol and water masers with which they are associated. This large survey was undertaken by an international team of 22 astronomers from 12 countries using NRAO's Very Long Baseline Array (VLBA) and achieved parallax accuracies of order $\pm$ 10 $\mu$as \citep{reid_etal_14}. The survey lasted about 15 years, starting with a pilot VLBI parallax measurement of W3(OH) in the Perseus arm of the Milky Way \citep{xu_etal_06a}. They collected candidate water and methanol masers with flux densities above 1 Jy from existing interstellar maser catalogs \citep{brand_etal_94, fontan_etal_10, caswel_etal_10, caswel_etal_11}. Additional surveys of maser candidates were carried out by the project team, particularly in the outer Galaxy where masers are sparse and distant. For maser candidates with uncertain absolute positions, VLA snapshot images were made to locate them to better than 0.1 arcsec \citep{hu_etal_16}.  They searched near these masers for extragalactic background sources of milliarcsecond size, to serve as fixed reference points: most of these came from existing calibrator catalogs \citep{petrov_etal_08, petrov_etal_11}. To increase the number of useful background sources in certain sky regions, they investigated the small-scale structures of more than 2000 potential reference sources from the NVSS and CORNISH surveys. Through more VLA snapshots and a VLBA survey they identified additional suitable background objects \citep{xu_etal_06b, immer_etal_11}. In total about 200 maser sources associated with HMSFRs in the bar and spiral arms of the Milky Way have had their trigonometric parallax measured in the BeSSeL project \citep{reid_etal_19}.

\begin{figure}[h!]
\centering
\includegraphics[width=0.9\textwidth, angle=0]{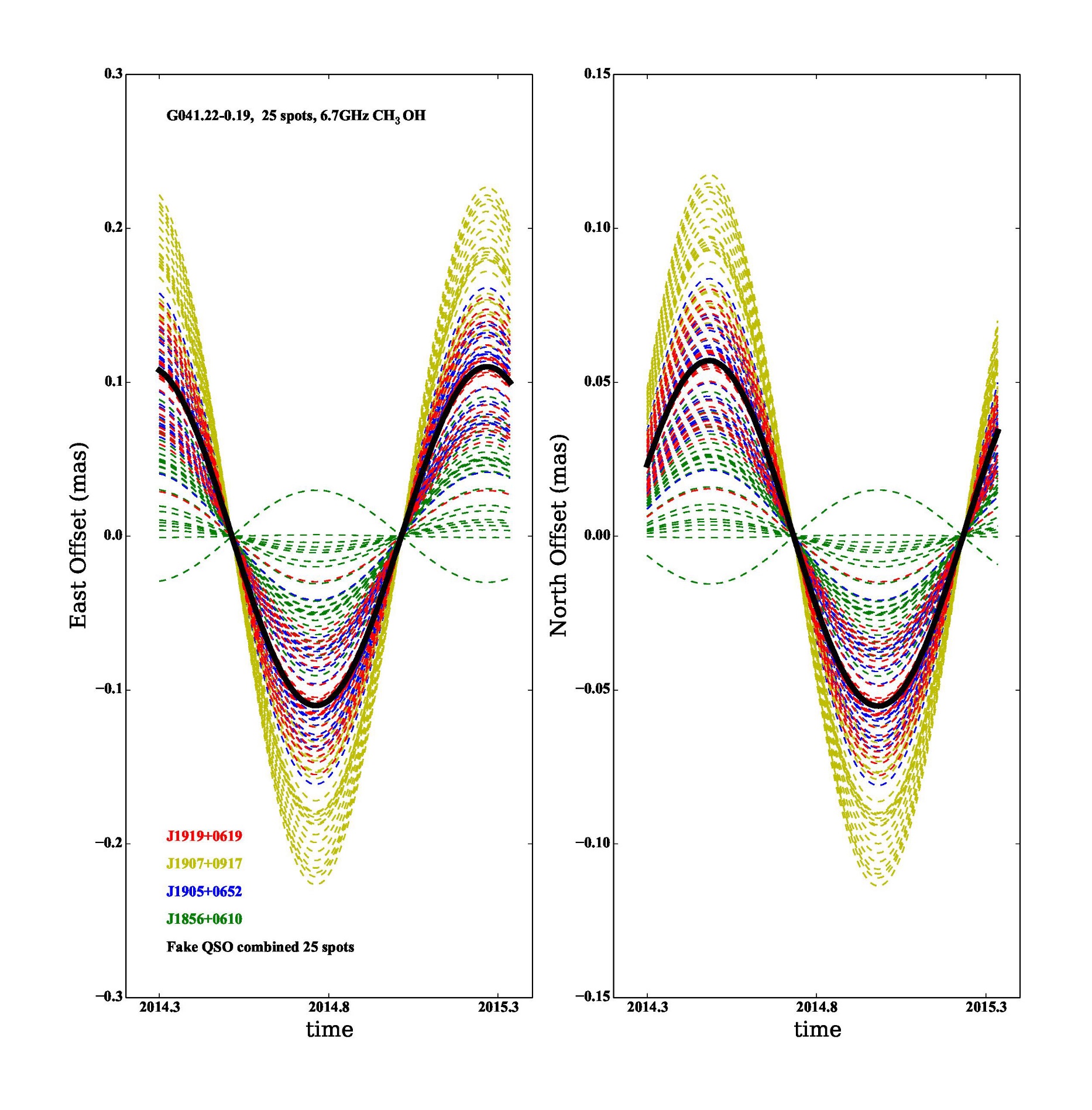}
\caption{The left (right) panel shows the East (North) parallax signature, with proper motion removed, for 6.7-GHz methanol maser G041.22-0.19 which is characterized by 4 background quasars and 25 maser spots, for 100 quasar-maser pairs. The final parallax, 0.113 $\pm$ 0.022 mas, is determined with the ‘fake quasar’ method \citep{reid_etal_17} and is shown as the thick solid line. Red, yellow, blue and green dashed lines denote individual fitting results for the four background quasars \citep{wu_etal_19}. }
\label{fig:parallax}
\end{figure}

%\Comment{Demonstration of parallax measurements with masers.}
%Figure 1 shows the results of VLBA observations of 22-GHz $H_2O$ masers associated with the HMSFR G074.03-01.71. The parallax of 0.629 $\pm$ 0.017 mas corresponds %to distance 1.59 kpc with accuracy 3\%, the best parallax accuracy in the BeSSeL project \citep{xu_etal_13}.
Distance uncertainties increase linearly with distance.  For most of the masers in the project the parallax measurements have less than 10\% error, owing to multiple spots in the maser sources and multiple background sources. For parallax measurements of methanol masers, many spots in the maser sources and more than one background sources near the HMSFRs have been found. G041.22-0.19, for example, consists of 25 maser spots and has four nearby background sources. There are 100 fitting parallaxes for the HMSFR \citep{wu_etal_19} as shown in Figure~\ref{fig:parallax}. This improves the accuracy of the parallax measurement but also brings complexity. A unique ‘fake quasar’ method to overcome the problem of signal propagation through the ionosphere from different background sources have developed \citep{reid_etal_17}. \citet{sanna_etal_17} have exceptionally measured the parallax of HMSFR G007.47+00.05 as 0.049 $\pm$ 0.006 mas, corresponding to a distance of 20.4 kpc with accuracy 12\%. The greatest distance measurements in the BeSSeL project are crucial to define the farthest segment of the Scutum-Centaurus spiral arm in the Milky Way.

%\Comment{Longer distance $\Rightarrow$ better coverage $\Rightarrow$ 4-arm spiral pattern emerges.}
Combining observations from the Japanese VERA project and the European VLBI Network, 199 trigonometric parallaxes and proper motions for water and methanol masers associated with HMSFRs have been employed to map the Milky Way, covering an area in the Galactic longitude from -2$^\degree$ to 240$^\degree$. Fitting log-periodic spirals to the locations of the masers and using well-established tangencies in the $4^{th}$ quadrant of the Galactic plane (Appendix in \citealt{reid_etal_19}), a new model of the spiral arm structure in the Milky Way has been established. It clearly shows four arms with some extra arm segments and spurs, as well as the Galactic bar and 3-kpc arm.
Considering the compositions in spiral arms of external spiral galaxies, such as giant HII regions, young OB stars, and young star clusters, filaments, the conceptual image of the Milky Way have been built and shown in the top panel of Figure~\ref{fig:complete}. It is currently the most scientifically accurate visualization of what the Milky Way looks like.

\subsection{Major Arms}

\begin{figure}
\centering
\includegraphics[width=0.8\textwidth, angle=0]{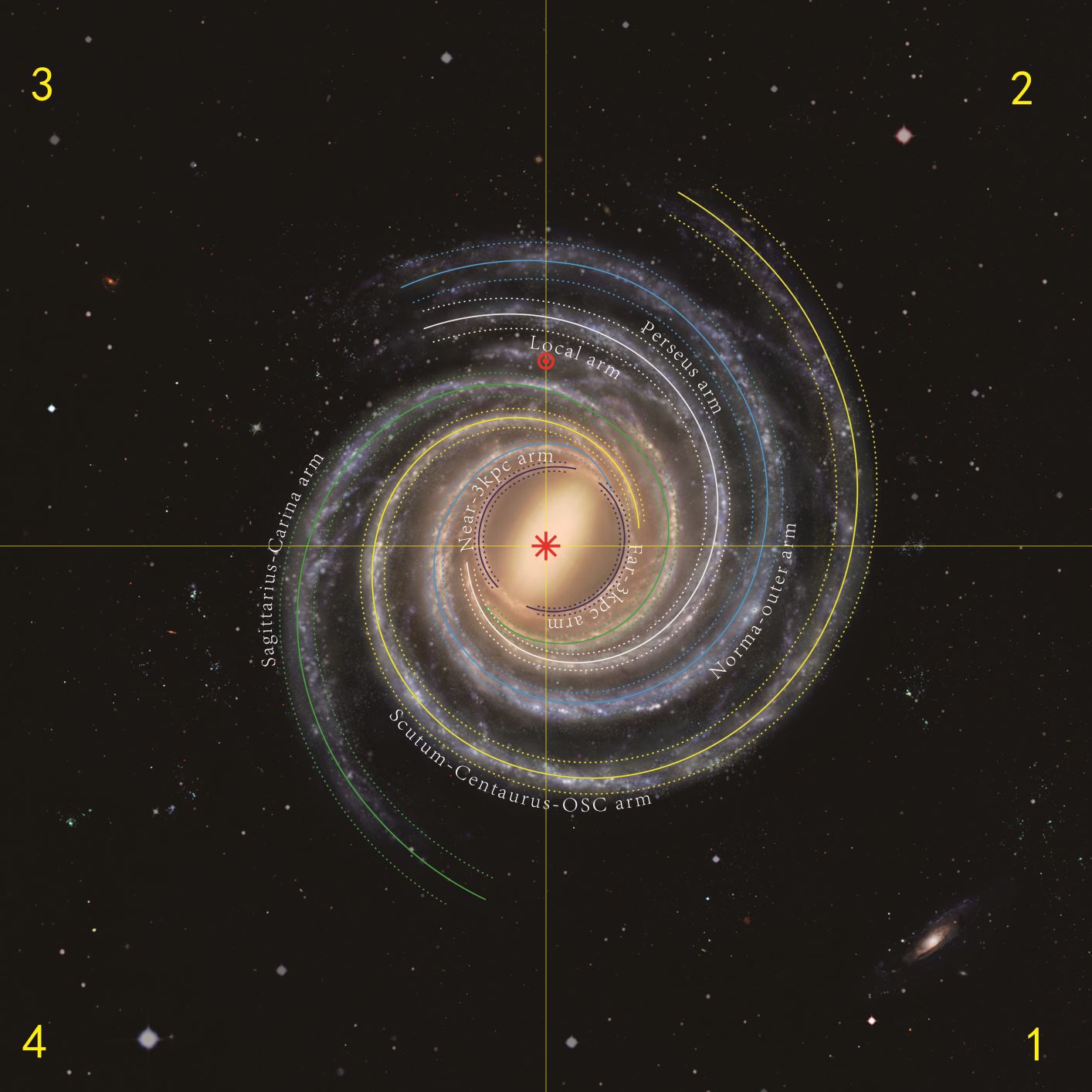}
\caption{The conceptual image overlaying the outlines of the four spiral arms of the Milky Way. The Galactocentric coordinate system is divided into four quadrants, as indicated by yellow dashed lines. Quadrant numbers are indicated in four corners. The Galactic center (red asterisk) is at (0, 0) and the Sun (red Sun symbol) is at (0, 8.15) kpc. The outlines of four arms are Norma-Outer arm (blue); Scutum-Centaurus-OSC arm (yellow); Sagittarius-Carina arm (green); and Perseus arm (white). The dotted lines are the widths of the four arms, defined as enclosing 90\% of their distance indicators \citep{reid_etal_19}.}
\label{fig:conceptual}
\end{figure}

%\Comment{Description of the updated schematic figure.}
Figure~\ref{fig:conceptual} shows a diagram overlaying the outlines of the four spiral arms on the conceptual image of the Milky Way. To represent the structure of the Milky Way, we have used the Galactocentric coordinate system with four quadrants. The Galactocentric azimuth $\beta$ is defined as 0 towards the Sun and increases clockwise. The elliptical bar in the center of the Milky Way extends from its nearer end in quadrant 2 into quadrant 4. Using red clump giant (RCG) star surveys, \citet{wegg_etal_15} found that the bar has semi-major and semi-minor axes of about 5 and 1.5 kpc and is oriented at about 30$^{\circ}$ to the line of sight \citep[see also][]{ratten_etal_07b,cao_etal_13}. The 3-kpc arm appears as a ring around the central bar \citep{vanwoe_etal_57, dame_etal_08}. The Norma-outer and the Scutum-Centaurus-OSC (Outer Scutum Centaurus) arms appear to start from the near end of the bar. The Norma-outer arm lies inside the Scutum-Centaurus-OSC arm. The Norma arm starts from near the end of the bar at $(x,\;y)$ = (2,\;3) kpc and extends into the $3^{rd}$ quadrant, passing counterclockwise through the $4^{th}$ quadrant before wrapping around the far end of the bar and becoming the Outer arm in the $1^{st}$ quadrant. The Scutum arm originates near $\beta$ = 90$^{\circ}$ in the $2^{nd}$ quadrant, and winds counterclockwise into the $3^{rd}$ quadrant as the Centaurus arm, which extends into the $1^{st}$ quadrant where it becomes the OSC arm. The Sagittarius-Carina arm and the Perseus arm both begin close to the far end of the bar, with the Sagittarius-Carina arm lying within the Perseus arm. The Sagittarius arm passes through the $4^{th}$, $1^{st}$ and $2^{nd}$ quadrants and becomes the Carina arm in the $3^{rd}$ quadrant before terminating in the 4th quadrant. The Perseus arm winds through the $4^{th}$, $1^{st}$ and $2^{nd}$ quadrants and appears to stop in the $3^{rd}$ quadrant. There are many spurs between the main arms in this picture of the Milky Way. Such spurs are common in spiral galaxies. Recently, \citet{ragan_etal_14} proposed that the spurs in the Milky Way may be giant molecular filaments. It is worth noting that the Spitzer/GLIMPSE survey using mid-infrared star counts \citep{benjam_etal_05} reported only two major stellar arms, the Perseus arm and the Scutum-Centaurus arm. A plausible reason to explain this discrepancy from BeSSeL could be that the other two arms contain excess gas but few old stars to be detected as stellar enhancement in GLIMPSE.

\begin{figure}
\centering
\includegraphics[width=0.8\textwidth, angle=0]{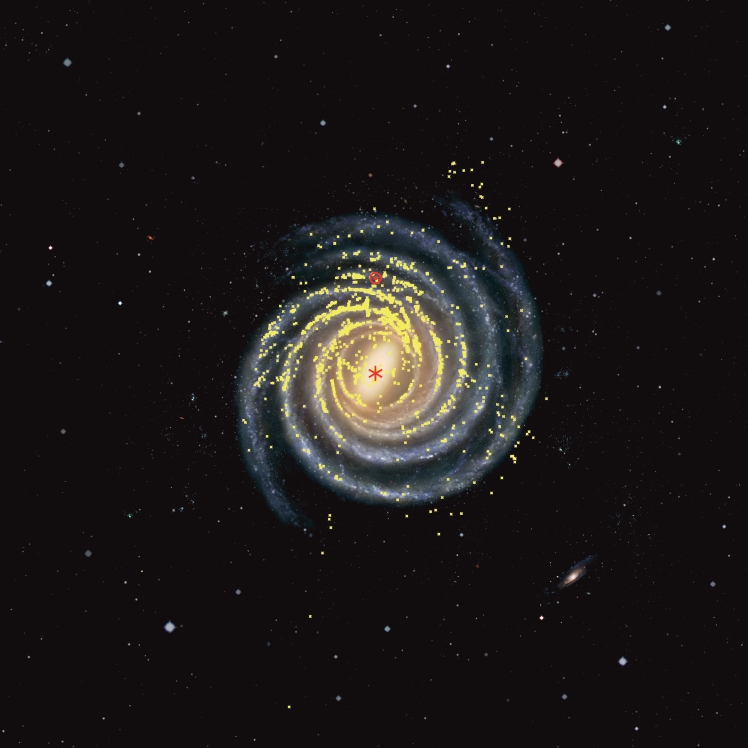}
\caption{A conceptual picture overlaid with the 2607 young objects \citep{dame_etal_01,valdet_etal_01, anders_etal_12,pestal_etal_05,sun_etal_15,green_etal_17} and 199 masers \citep{reid_etal_19} shown as yellow points.  }
\label{fig:youngobjects}
\end{figure}

VLBI parallax and proper motion measurements use one of the most sophisticated instrumental system and the phase calibration in astronomy. Especially, there are strict restrictions on the observed targets which must be strong and compact.  Only about 300 masers in the HMSFRs have been measured with VLBI astrometry. Fortunately, there are various kinds of young objects associated the HMSFRs, such as HII regions, giant molecular clouds (GMCs), massive outflows as well as maser with weak intensity. \citet{reid_etal_17} have developed a Bayesian distance calculator and used longitude-latitude-velocity values of these young objects to re-estimate their distances, refining the standard kinematic values. 2607 young objects have been collected from several surveys \citep{dame_etal_01,valdet_etal_01, anders_etal_12,pestal_etal_05,sun_etal_15,green_etal_17}. We have overlaid these 2607 young objects and 199 masers by parallax measurement \citep{reid_etal_19} on the conceptual image of the Milky Way shown in Figure~\ref{fig:youngobjects}. It is obvious that young objects clearly delineate the spiral arms in the Milky Way.

\subsection{The Local Arm}
\label{sect:localarm}

%Especially the Local arm is illustrated as an isolated section because the Sun is closely located at. The structure and kinematics of the Local arm is relatively easy to investigate by stellar objects and gas around sun. It may provide significant clues about the origin of main arms.

Previously, the local arm was supposed to be a short fragment similar to smaller appendages seen branching off spiral arms in other galaxies and called the Orion or Local spur \citep{vandeh_etal_54, geo_geo_76}. Recently, more than 30 methanol (6.7-GHz) and water (22-GHz) masers in high-mass star-forming regions around the Sun have been measured their parallax and proper motions with the distance accuracy of better than $\pm$10\% and even 3\%, the best parallax measurement in the BeSSeL project. The accurate locations of interstellar masers in HMSFRs have been shown that the Local arm appears to be an orphan segment between the Sagittarius and Perseus arms that wraps around less than a quarter of the Milky Way. The segment has a length of $\sim 6\; \kpc$ and the width of $\sim 1 \;\kpc$ with a pitch angle from 10.1$^{\circ} \pm$ 2.7$^{\circ}$ to 11.6$^{\circ} \pm$  1.8$^{\circ}$. These results reveal that the Local arm is larger than previously thought, and both its pitch angle and star formation rate are comparable to those of the Galaxy's major spiral arms. The Local arm is reasonably referred to as the fifth feature in the Milky Way. The ``spur'' interpretation is definitely incorrect \citep{xu_etal_13,xu_etal_16,reid_etal_19,hirota_etal_20}.

To understand the form of the Local arm between the Sagittarius and Perseus arms, the stellar density of a specific population of stars with about 1 Gyr of age between 90$^{\circ} \leq l \leq 270^{\circ}$ have been mapped using the {\it Gaia} DR2 \citep{miyach_etal_19}. The 1 Gyr population have been employed because they are significantly evolved objects than the gas in HMSFRs tracing the Local arm. \citet{miyach_etal_19}  have carried out an interesting investigation to compare both the stellar density and gas distribution along the Local arm.  They found a marginally significant arm-like stellar overdensity close to the Local arm, identified with the HMSFRs especially in the region of 90$^{\circ} \leq l \leq 190^{\circ}$. They have concluded the Local arm as the arm segment associated with not only the gas and star-forming clouds, but also a significant stellar overdensity. Additionally they found that the pitch angle of the stellar arm is slightly larger than the gas-defined arm, and also there is an offset between HMSFR-defined and stellar arm. The offset and different pitch angles between the stellar and HMSFR-defined spiral arms are consistent with the expectation that star formation lags the gas compression in a spiral density wave lasting longer than the typical star formation timescale of $\sim 10^7-10^8$ years.
%The results may popularize the origin of four main arms in the Milky Way.

\section{Summary and Outlook}
\label{sect:summary}

Our home Galaxy is the closest galaxy that we can study in exquisite detail. Yet understanding the structure and kinematics of our Milky Way bulge is not a trivial task, mostly because of our disadvantageous vantage point in the disk and severe dust extinction. Despite these challenges we have made giant leaps in understanding the Galactic structure in the past decade. The BeSSeL survey has drawn a picture of the most reliable spiral arm structure of the Milky Way to date. Careful analysis and modelling of extensive datasets on the inner Galaxy reveal more details of the Milky Way bar/bulge. To our current best knowledge, the Milky Way contains a long strong bar, four major spiral arms, and a local arm that may be longer than previously thought. The Galactic bar transitions smoothly from a central peanut-shaped structure to an extended thin part that ends around $R\sim 5 \;\kpc$. Most of the boxy Galactic bulge that we observe is probably just the centrally thickened, peanut-shaped bar viewed nearly end-on.

Although we have some new findings, we are also left with many more unanswered questions and puzzles. About two thirds of spiral galaxies are barred, so in this aspect our barred Milky Way is certainly in the majority. Our Galaxy possesses four clearly-defined major spiral arms and an additional smaller Local arm, which are not  the most common form of spiral structure. In this sense our Milky Way is probably a normal, but not a typical spiral galaxy.

% Competing theories on the origin of spiral structure have been posed.
The origin of grand-design spirals in galaxies is still actively debated in the community, and the Galaxy's configuration of the four plus one arms may pose an even bigger challenge for theorists. The standard Lin-Shu quasi-steady theory argues that gravitational instabilities on the scale of the entire galaxy form grand-design spiral-wave patterns lasting for almost the lifetime of a galaxy. Other theories propose that more short-lived spirals re-emerge many times over billions of years, but the detailed mechanism, such as a recurrent cycle of groove modes \citep{sel_car_19}, or tidally induced, is still not completely settled. More comprehensive reviews on spiral structure may be found in \citet{sellwood_14}, \citet{shu_16}, and \citet{sel_car_19}. The next generation of radio telescope arrays capable of VLBI, such as the Square Kilometer Array in South Africa/Australia and the Next Generation Very Large Array in North America, will detect even fainter  radio emissions from much more distant stars. These planned arrays will map out more accurately the large-scale spiral structure and the interface region where the bar ends in unprecedented detail. The more accurate characterization of spiral structure, coupled with the improved phase space structures by future {\it Gaia} releases, may hopefully provide more clues to distinguish competing theories of how the Galaxy's spiral structure formed \citep[e.g.][]{sel_tri_19}.

The Galactic bar/bulge also contains crucial information about the formation of evolutionary history of the Milky Way. There are still many open questions to be answered in more sophisticated chemo-dynamical models of better data. For example, how many distinct metallicity components are there in the Galactic bulge? How do they vary spatially, and how do they correlate with kinematics? 
%How is the fossil record of the early inner Galaxy (thick disk, old and younger thin disks) mapped into the bulge structure?
How is the early inner Galaxy (thick disk, old and younger thin disks) gradually mapped into the presently-observed bulge structure?
How does the outer bulge/bar transition into the inner Galactic halo? Ongoing and upcoming large surveys promise to shed new light on these questions about the Milky Way bar/bulge. Parallaxes and proper motions of about 20 million bulge stars \citep{robin_etal_05} will be further improved by future \textit{Gaia} data releases. The Blanco Dark Energy Camera (DECam) Bulge survey is a Vera Rubin Observatory (LSST) pathfinder imaging survey  of the relatively less reddened Galactic bulge \citep{rich_etal_20,johnso_etal_20}.  Optical photometry in SDSS $u$  + Pan-STARRS $grizy$ bands can provide a large color baseline to investigate the age and metallicity distributions of the major structures of the Galactic bulge. These data will be more powerful when combined with other surveys such as APOGEE, VVV, \textit{Gaia}-ESO, and GIBS. With the large influx of data and the improvement in more sophisticated chemo-dynamical models, greater progress is expected in putting together all the pieces of the Milky Way bar/bulge puzzle.

\begin{acknowledgements}
We thank Shude Mao, Zhi Li, Zhao-Yu Li, and Iulia Simion for helpful discussions and comments. The research presented here is partially supported by the National Key R\&D Program of China under grant No. 2018YFA0404501; by the National Natural Science Foundation of China under grant Nos. 11773052, 11761131016, 11333003; by the ``111'' Project of the Ministry of Education under grant No. B20019, and by the MOE Key Lab for Particle Physics, Astrophysics and Cosmology. This work made use of the Gravity Supercomputer at the Department of Astronomy, Shanghai Jiao Tong University, and the facilities of the Center for High Performance Computing at Shanghai Astronomical Observatory.

X-W Zheng gratefully acknowledges support from National Natural Science Foundation of China (under grants Nos. 10073004, 19673006, 10133020, 10673024, 11073054, 1113308) and the Research Priority Program of Nanjing University and help from Z-Q Zhu (Nanjing University of Arts), P-F Chen (Nanjing University), and X-B-N Ji (Nanjing University).
\end{acknowledgements}

\bibliographystyle{raa}
\newcommand{\noopsort}[1]{} \newcommand{\singleletter}[1]{#1}

%\begin{figure}
%   \centering
%   \includegraphics[width=8.0cm, angle=0]{ms0136fig1.eps}
%  % \begin{minipage}[]{85mm}
%   \caption{The temperature as a function of height above the photosphere. Several lines are overplotted on the curve based on their formation temperatures. }
%%\end{minipage}
%   \label{Fig1}
%   \end{figure}

%
%\begin{thebibliography}{99}
%\small \setlength{\itemindent}{-3mm}
%\setlength{\itemsep}{-0.5mm}
%\setlength{\baselineskip}{4.7mm}
%
%\bibitem[{Abdo} {et~al.}(2010)]{A10}
%{Abdo}, A.~A., {Ackermann}, M., {Ajello}, M., {et~al.} 2010, \apj, 723, 1082

%\end{thebibliography}

%\bibliographystyle{/home/lixh/raa/bibstyle/raa}
%\bibliography{bibtex}

\end{document}